\documentclass[nofootinbib,showpacs,preprintnumbers,amsmath,amssymb,floatfix]
{revtex4-2}
\usepackage{graphicx}
\usepackage{wrapfig}
\usepackage{epstopdf}

\begin{document}


\title{Comment on applying dispersion relations to amplitudes with infrared singularities}

\vspace*{0.3 cm}

\author{B.I.~Ermolaev}
\affiliation{Ioffe  Institute, 194021
 St.Petersburg, Russia}
\author{S.I.~Troyan}
\affiliation{St.Petersburg Institute of Nuclear Physics, 188300
Gatchina, Russia}

\begin{abstract}
Dispersion Relations (DR) are known to be a powerful instrument for studying scattering amplitudes. In particular, they often apply to calculations in Perturbative QCD and Standard Model. We argue that applying DR to amplitudes with double-logarithmic (DL) contributions should be done with a proper caution because DL terms are often infrared-divergent. Ignoring  this circumstance leads to incorrect results.
As an example of such situation, we consider
applying DR to decays of on-shell
$W$ and $Z$ -bosons into fermion pairs in the Double-Logarithmic Approximation
\end{abstract}

\pacs{12.38.Cy}

\maketitle

\section{Introduction}

Dispersion Relations (DR)  are often used in Perturbative QCD and Standard Model to simplify calculations of various reactions at high energies both in fixed order calculations and in approaches involving total
resummations.  In particular, they often apply to calculate scattering amplitudes in Double-Logarithmic Approximation (DLA). In this case, calculation of amplitude $M (s)$, with $s$ being the invariant total energy,
consists of two steps: First, calculation of $\Im M$, and then calculation of $M$, using DR with one substraction:

\begin{equation}\label{dr}
	M(s) = \frac{s}{\pi} \int_{s_0}^{\infty} d s' \frac{\Im M (s')}{s' (s' - s)}  \approx -
	\frac{1}{\pi} \int_{s_0}^s d s' \frac{\Im M (s')}{s'}~.
\end{equation}

Applying DR considerably simplifies
calculations because  \\
\textbf{(i)} Imaginary parts are free of the ultraviolet divergences, so the use of
DR makes possible avoiding the procedure of renormalization.\\
\textbf{(ii)} Calculating $\Im M$ is simpler technically because the cut propagators therein
are replaced by the delta-functions.  \\
 \textbf{(iii)}
The graphs with zero imaginary parts can be neglected at once. \\
The latter item was  frequently used by L.N.~Lipatov to calculations in the Leading logarithmic Approximation, see e.g. Ref.~\cite{ln}. \emph{}
However, we demonstrate in the present paper that applying it to calculations in DLA should be done with a proper caution to avoid incorrect conclusions.   To illustrate our point, we consider very simple examples. Namely, amplitudes of decays of electroweak bosons in fermion pairs. First, decay of $Z$-boson and then decay of $W$-boson.
We demonstrate that total resummation of radiative corrections to the Born amplitudes
in DLA leads to Sudakov-like expressions for those amplitudes and explain which mistakes could be made, when Argument
(\textbf{iii}) is applied in an inappropriate way.\\

Our paper is organized as follows: in Sect.~II we remind the basic features of the Sudakov form factor, which we will use in the paper. Sect.~III is for considering decay of the on-shell $Z$-boson in DLA. Decay of the on-shell $W$-boson is considered in Sect.~IV. Finally, Sect.~V is for concluding remarks.

\section{Basic formulae for the Sudakov form factor}

Let us consider the vertex $\Gamma_{\lambda}$ of the electromagnetic interaction of fermions. It includes two form factors,
$f(q^2)$ and $g(q^2)$:
\begin{equation}\label{emgamma}
\Gamma_{\lambda} = \bar{u} (p_2) \left[ \gamma_{\lambda} f(q^2) -
\frac{\sigma_{\lambda\nu} q_{\nu}}{2 m}g(q^2)\right] u(p_1) ~,
\end{equation}
where $q$ is momentum of the virtual photon and $p_{1,2}$ are the fermion momenta, so that $q = p_2 - p_1$. When $q^2 < 0$, Eq.~(\ref{emgamma}) describes scattering of
a virtual photon $\gamma^*$
with momentum $q$ off a fermion with momentum $p_1$. When $q^2 > 0$, $\Gamma_{\lambda}$ describes decays of a virtual photon $\gamma^*$ into fermion-antifermion pairs
$f \bar{f}$:

\begin{equation}\label{photondec}
\gamma^* (q) \to f(p_1) \bar{f} (p_2),
\end{equation}
where $q = p_2 + p_1$, so $q^2$ is positive.
Behavior of the form-factor $f(q^2)$ at negative and large $q^2$, i.e. at

\begin{equation}\label{qbig}
|q^2| \gg  |p^2_{1,2}|,
\end{equation}
was
considered first in  Ref.~\cite{sud}. This resulted in the  discovery of  the double-logarithmic (DL)
contributions in QED. Total resummation of DL terms leads to the famous Sudakov exponents which rapidly fall at
both positive and negative $q^2$ providing $|q^2|$ is large.

Now we focus on the decay channel in Eq.~(\ref{photondec}).
 The fermions in Eq.~(\ref{photondec}) can be either off-shell or on-shell. The form factors for these situations are different. Keeping
 $f(q^2)$ as a generic notation, we introduce the notation
 $f^{\prime}(q^2)$ for the form factor in the kinematics where the produced  fermions are off-shell, with $|p^2_{1,2}| \gg m^2$ ($m$ is the fermion mass),
  and denote the form factor $f^{\prime \prime}(q^2)$ when the fermions are on-shell.  The form factor $f^{\prime}(q^2)$ is IR-stable:

\begin{equation}\label{sudoff}
	f^{\prime}(q^2) = \exp \left[- \frac{\alpha}{2 \pi}L^{\prime}\right],
\end{equation}
with

\begin{equation}\label{loff}
L^{\prime} = \ln \left(-q^2/|p^2_1|\right)
	\ln \left(-q^2/|p^2_2| \right).
\end{equation}

On the contrary, the form factor $f^{\prime \prime}(q^2)$ is IR-divergent, so
an IR cut-off is needed. As a result, Eq.~(\ref{sudoff}) is replaced by

\begin{equation}\label{sudon}
	f^{\prime \prime}(q^2) = \exp \left[- \frac{\alpha}{4 \pi} L^{\prime \prime} \right],
\end{equation}
where
\begin{equation}\label{lon}
L^{\prime \prime} =  \ln^2 (q^2/\mu^2) - \Theta (m^2 - \mu^2)  \ln^2(m^2/\mu^2),
\end{equation}
with  $\mu$ being the infrared (IR) cut-off.
The exponential form of  $f(q^2)$ was also obtained
in  the QCD context in Ref.~\cite{quarkff} and in
Standard Model in Ref.~\cite{flmm}. A similar exponentiation of
DL contributions to the form factor $g(q^2)$ of Eq.~(\ref{emgamma})
was obtained in Ref.~\cite{et}, which finalized studying the electromagnetic vertex $\Gamma_{\lambda}$ in DLA:

\begin{equation}\label{gamma}
\Gamma_{\lambda} = \bar{u} (p_2) \left[ \gamma_{\lambda} f(q^2) +
\frac{\sigma_{\lambda\nu} q_{\nu}}{ m} ~\frac{d f(q^2)}{d \rho}\right] u(p_1),
\end{equation}
where $\rho = \ln (-q^2/\mu^2)$ and $f(q^2)$ can stand for either the off-shell or on-shell form factor.
Denoting electric charges of $e^+$ and $e^-$ as $e \bar{Q}$ and $e Q$ respectively, with $Q = -1$ and $\bar{Q} = - Q =  +1$, we can write Eqs.~(\ref{sudoff},\ref{sudon}) at positive $q^2$ in the following form:

\begin{eqnarray}\label{aprodq}
f^{\prime \prime} &=&  \exp \left[\frac{\alpha}{4 \pi} \bar{Q} Q L^{\prime \prime} \right] = \exp \left[-\frac{\alpha}{4 \pi}Q^2L^{\prime \prime} \right]
 = \exp \left[- \frac{\alpha}{4 \pi} \bar{Q}^2L^{\prime \prime} \right],
 \\ \nonumber
 f^{\prime} &=&  \exp \left[\frac{\alpha}{2 \pi} \bar{Q} Q L^{ \prime} \right] = \exp \left[-\frac{\alpha}{2 \pi}Q^2L^{\prime} \right]
 = \exp \left[- \frac{\alpha}{2 \pi} \bar{Q}^2L^{\prime} \right].
\end{eqnarray}

Relations between on-shell and off-shell scattering amplitudes in DLA were considered in detail in Ref.~\cite{egot}.

\section{DL corrections to the decay $Z \to f \bar{f}$}

In this section we consider the decay of $Z$ -boson into the fermion-antifermion pair $f \bar{f}$:

\begin{equation}\label{zdecay}
	Z(q)   \to f (p_1) \bar{f} (p_2),
\end{equation}
presuming that both
$Z$ -boson and produced fermions are on-shell and   the fermion mass $m$ is small compared to the mass $m_Z$
of $Z$-boson: $m_Z \gg m$, which makes reasonable calculating this graph in DLA. We also presume that the produced fermions have non-zero electric charges and
denote electric charges of the produced fermion and antifermion  $Q_f$ and $\bar{Q}_f$ respectively. The first-loop contribution, $A_Z^{(1)}$
corresponds to the graph in Fig.~1.
\begin{figure}
\includegraphics[width=.8\textwidth]{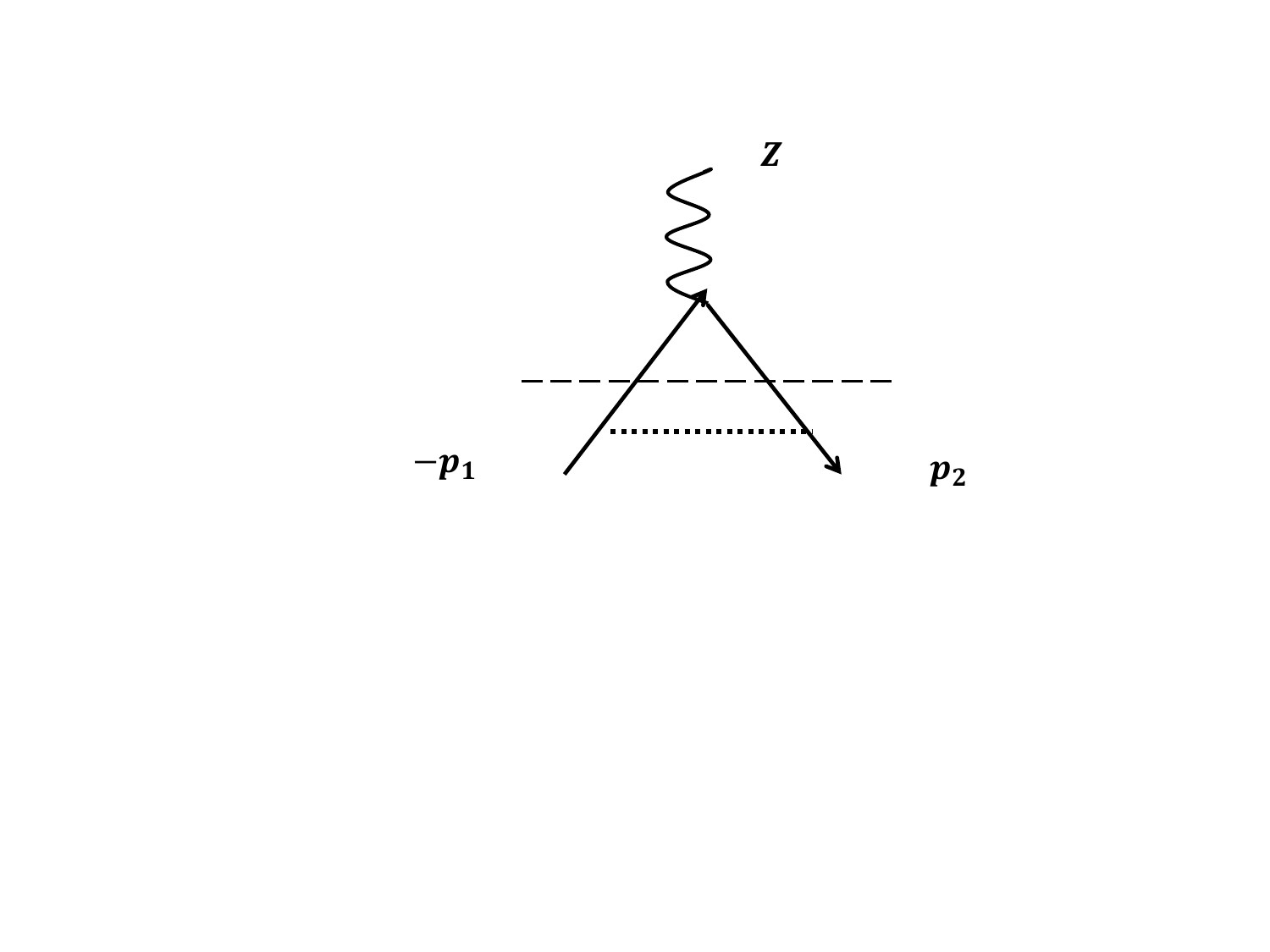}
\caption{\label{DRfig1} First-loop contribution to $Z \to f \bar{f}$. The waved line is for the $Z$-boson, straight lines denote fermions and
the dotted line is for the virtual photon. The dashed line denotes the cut. }
\end{figure}
Applying the Dispersion Relations to this graph and using Eq.~(\ref{aprodq}),
we obtain

\begin{equation}\label{az1}
A_Z^{(1)} = A_Z^{Born} \left[\frac{\alpha Q_f \bar{Q}_f}{4 \pi} L_Z  \right],
\end{equation}
with $L_Z$ defined as follows:

\begin{equation}\label{lz}
L_Z =  \ln^2 (m_Z^2/\mu^2) - \Theta (m^2_f - \mu^2)  \ln^2(m^2_f/\mu^2).
\end{equation}

Exponentiation of $A_Z^{(1)}$ allows us to obtain the expression for $A_Z$ in DLA:

\begin{equation}\label{az}
A_Z = A_Z^{Born} \exp \left[ \frac{\alpha Q_f \bar{Q}_f}{4 \pi} L_Z\right] =
A_Z^{Born} \exp \left[- \frac{\alpha Q_f^2}{4 \pi} L_Z\right] ~.
\end{equation}

The exponent in Eq.~(\ref{az}) has the negative sign, so accounting for DL corrections suppresses $A^{Born}_Z$,
 which is typical for  logarithms of the Sudakov type.

\section{DL corrections to the decay $W \to f' \bar{f}$}

Now consider decay of the on-shell $W$-boson into the on-shell fermion couple:

\begin{equation}\label{wdecay}
	W \to f^{\prime}(p_1) \bar{f}(p_2).
\end{equation}

To begin with, consider this process in the first loop. Denote the electric charges of the fermion (antifermion) $Q^{\prime}_f$ ($\bar{Q}_f$)
and denote their masses $m^{\prime}_f$ and $m_f$  respectively. We also
assume for simplicity that $Q^{\prime}_f \neq 0$ and $\bar{Q}_f \neq 0$, though  the final answer will not be sensitive to this assumption.
The first-loop calculation involves three graphs depicted in Fig.~2.

\begin{figure}
\includegraphics[width=.8\textwidth]{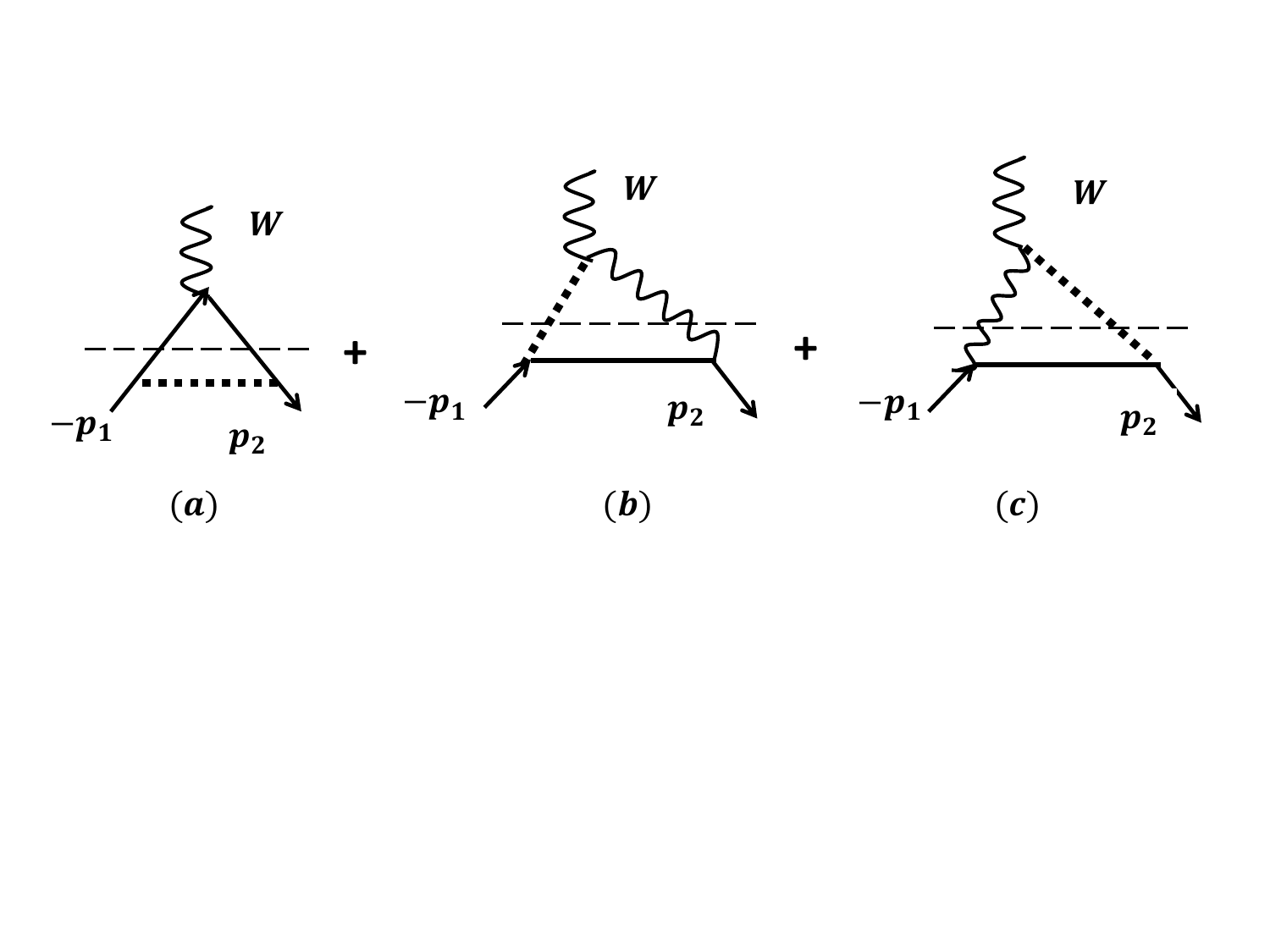}
\caption{\label{DRfig2} First-loop contributions to the amplitude of $W$-decay into fermions. The waved lines are for the $W$-boson, straight lines denote the fermions and
the dotted lines are for the virtual photons. The dashed lines denote the cuts. }
\end{figure}

Dealing with graph (2a) is quite similar to the calculation in the previous Sect save the fact that the masses of the fermion and antifermion can differ from each other.  It leads to the following result:
\begin{equation}\label{aw2a}
A^{(2a)}_W = A_W^{Born} \left[\frac{\alpha Q'_f \bar{Q}_f}{4 \pi} L_W \right]~,
\end{equation}
where

\begin{equation}\label{lw}
L_W =  \ln^2 (m_W^2/\mu^2) - \frac{1}{2}\Theta ({m^{\prime}}^2_f - \mu^2)  \ln^2({m^{\prime}}^2_f/\mu^2) - \frac{1}{2}\Theta (m^2_f - \mu^2)  \ln^2(m^2_f/\mu^2).
\end{equation}

Let us notice that $Q'_f \bar{Q}_f > 0$, so
had the graph (2a) been the only first-loop contribution to $A_W^{(1)}$, its exponentiation would lead to
enhancement of $A_W^{Born}$ by DL corrections instead of the standard Sudakov suppression as it was e.g. in Eq.~(\ref{az1}).

Now proceed to graphs (2b) and (2c) and denote  their DL contributions $A^{(2b)}_W$ and $A^{(2c)}_W$ respectively. Obviously, the amplitudes above the cuts in Figs.~(2b,2c) correspond to emission/absorption of on-shell photons,
with $k^2 = 0$,  by
on-shell $W$-bosons. Such processes are forbidden by the momentum conservation and therefore  $A^{(2b)}_W = A^{(2c)}_W = 0$.
However, this argumentation is not quite correct: the point is that the photons in the expressions corresponding to graphs (2b,2c)  cannot be on-shell because
these expressions are singular at  $k^2 = 0$. Such infrared (IR)
singularities should be regulated with introducing an IR cut-off, which prevents the
photons to be on-shell. It means that there is no violation of the momentum conservation and justifies applying DR to graphs (2b,2c).
As a result, we obtain

\begin{eqnarray}\label{aw2b2c}
A^{(2b)}_W &=& A^{Born}_W \left[-\frac{\alpha Q'_f Q_W}{8 \pi} L_W \right],
\\ \nonumber
A^{(2c)}_W &=& A^{Born}_W \left[-\frac{\alpha \bar{Q}_f Q_W}{8 \pi} L_W \right],
\end{eqnarray}
where $Q_W$ is the electric charge of the $W$-boson.
Adding up $A^{(2a)}_W, A^{(2b)}_W $ and $A^{(2c)}_W $, and using that $Q_W = Q'_f +  \bar{Q}_f$, we arrive at the first-loop contribution $A^{(1)}_W $:

\begin{equation}\label{aw1}
A^{(1)}_W = A^{Born}_W \left[-\frac{\alpha}{8 \pi} \left({Q'_f}^2 +  {\bar{Q}_f}^2 \right) L_W \right].
\end{equation}

Its exponentiation yields amplitude $A_W$ in DLA:

\begin{equation}\label{aw}
A_W = A^{Born}_W \exp \left[-\frac{\alpha}{8 \pi} \left({Q'_f}^2 +  {\bar{Q}_f}^2 \right) L_W \right].
\end{equation}

The amplitudes $A^{(1)}_W$ and $A_W$ in Eqs.~(\ref{aw1},\ref{aw}) are represented in the form where either $Q'_f$ or $\bar{Q}_f$
can be equal to zero. It makes possible to apply these formulae to the $W$-decays with participation of neutrinos, i.e.
$W \to l\bar{\nu}, W \to l \nu$, with $l$ denoting leptons.

\section{Conclusion}
In the present paper we have examined
practical use of the Dispersions Relations. There is
the very important technical issue in this subject:
the Feynman graphs with zero imaginary parts can be left out.
Admitting correctness of this statement in general, we argued for its careful application.
We have shown that there are cases when the imaginary parts (which formally should have
vanished because of the momentum conservation)
do not vanish in reality. In particular, it applies to the graphs, whose  imaginary parts contain
DL contributions induced by soft photons. When such photons are put on their mass shell,
the DL contributions
become IR-divergent. Introducing the IR cut-off  to regulate the
divergences
provides the photons with a fictitious mass, which blocks vanishing the imaginary parts.
This observation should be taken into account before neglecting the graphs with zero imaginary parts, when Dispersion Relations are used.

%

\end{document}